# Negative Probability Sampling in Study of Reflection Surface Electron Spectroscopy Spectrum

B. Da[1,2], S. F. Mao[3], Z. J. Ding[1]

[1]*Hefei National Laboratory for Physical Sciences at Microscale and Department of Physics, University of Science and Technology of China, Hefei, Anhui 230026, P.R. China*

[2]*International Center for Young Scientists (ICYS), National Institute for Materials Science (NIMS), 1-2-1 Sengen, Tsukuba, Ibaraki 305-0044, Japan*

[3]*School of Nuclear Science and Technology, University of Science and Technology of China, Hefei, Anhui 230026, P.R. China*

**ABSTRACT**

We propose a sampling method to include the negative contribution to probability density distribution in a sampling procedure. This sampling method is a universal solution for all negative probability problem and shows extraordinarily power in negative cross section problem. A Monte Carlo simulation including negative cross section contribution is developed and successfully preformed to simulate reflection electron energy loss spectroscopy (REELS) spectra for Ag and Au as examples. Excellent agreement is found between simulated spectra and experimental measurements. Notably improved fits to experimental REELS spectra in low energy loss range illustrate the method's strength as well as the necessity of negative cross section contribution.

**PACS numbers**: 34.50.Bw, 73.40.-c, 02.50.Cw

Probability theory is an important tool in modern physics. A great discovery of twentieth century physics was the probabilistic nature of physical phenomena at microscopic scales, described in quantum mechanics. At present there is a firm consensus among the physicists that probability theory is both essential and necessary for physicists to describe quantum phenomena. However, at the same time, problems of physics forced physicists to consider not only using traditionally probability but also negative probability.

Dirac [1] and Heisenberg [2] first introduced probability distributions with negative values into physics in 1930s within the context of quantum theory, however, missed the significance of negative values. Two years later, the Wigner quasi-probability distribution which can and normally does go negative for states which have no classical model was introduced by Eugene Wigner to study quantum corrections to classical statistical mechanics, in order to link the wave function that appears in Schrodinger's equation to a probability distribution in phase space. Dirac [4] not only supported Wigner`s approach but also gave a physical concept to the negative probabilities and energy. He thought that negative energies and probabilities are well-defined concepts mathematically and should not be considered as nonsense, like a negative of money and described other useful physical interpretations of negative probabilities. Pauli [5] also gave his opinion on the negative probabilities by a simple but intuitive example. In this example, Pauli demonstrated how the mathematical models work when the negative probabilities generated by the renormalization procedure.

Feynman [6] thought that the only difference between a probabilistic classical world and the equations of the quantum world is that somehow or other it appears as if the probabilities would have to go negative and then introduced the concept of negative probability in the context of Young's double-slit experiment and in doing so sheds a new light on the problem [7]. Later Feynman [8] wrote a special paper on negative probability where he discussed different examples demonstrating how negative probabilities naturally come to physics and beyond. Bayes formula for conditional probabilities were employed in this discussion as

$$P(i) = \sum_{\alpha} P(i|\alpha) P(\alpha), \qquad (1)$$

where $\sum_{\alpha} P(\alpha) = 1$. The idea is that as long as $P(i)$ is positive then it is not a problem if some of the probabilities $P(i|\alpha)$ or $P(\alpha)$ are negative or larger than unity. This approach works well when one cannot measure all of the conditional probabilities $P(i|\alpha)$ or the unconditional probabilities $P(\alpha)$ in an experiment.

Although such an approach has been used in quantum physic to solve conceptual problems [9-14] and shown its advantages in shedding a new light on these problems, however, there is still a blank in application of negative probability to solve problems never solved before. Furthermore, physicists still are helpless in the face of appearance of negative probability in some well-established theories due to the lacking of reliable sampling method. Therefore, a sampling method for negative probability is desperately needed.

In this work we propose a sampling method to deal with negative probability following the interpretations from Feynman [8]. In this method, negative probability is considered as reflecting the immediate interaction of probabilities, i.e. suppression to the positive probabilities and can be performed to a system by the changing in population of a system, positive for increase and negative for decrease. It is worth noting that the states of negative probability in whatever theories cannot happen, but they, where they are, decrease the sum probability of the integrally positive regions of the probability density distribution.

To deal with the negative probabilities in practical problem, a solid mathematical underpinning is particularly crucial. Fortunately, rigorous mathematical foundations for negative probability has been built by Burgin [15] in which negative probability was unified in the form of extended probability and a frequency interpretation of negative probability was given. Here we use a simple example to illustrate how our proposed sampling method works as shown in Fig. 1.

Let us consider a set $\Omega$ described by the probability density function $f(x)$, which consists of two irreducible parts $\Omega^+$ and $\Omega^-$ due the sign of $f(x)$, i.e. the probability density function of part $\Omega^+$ and $\Omega^-$ are, respectively, positive and negative according to the different determinate range of variable. Subsets from the set $\Omega^+$ are called provisional positive events, while subsets of the set $\Omega^-$ are called provisional negative events. These negative events are common in physic filed as an annihilation and usually connected to negative objects, such as encountering an antiparticle. Due to the different attitudes for these negative part $\Omega^-$, there are various possible treatments. The thoughtless one is employing the strict definition in conventional probability theory, this treatment is unacceptable because of the confusion in classifying the random event in the sampling procedure from the decreasing cumulative function. The most popular treatment is mandatorily neglecting these negative possibilities and simplify setting these values to zero, however, according to this approximation, these provisional negative events contributions are consciously ignored. In this work, we employed a variant of the definition of accumulative function $F(x) = \int_{-\infty}^{x} |f(x')| dx'$ to include the contributions from provisional negative events. Differing from the conventional sampling method, provisional negative events have possibility to be sampled with a uniform random number in this sampling procedure according to the proportion of negative events, i.e. $|\Omega^-|/(\Omega^+ + |\Omega^-|)$. After determining the quantitative information about these provisional negative events, we employed the essence of the "demon algorithm" method [16], a Monte Carlo method for efficiently sampling members of a micro canonical ensemble with a given energy, to take into account these provisional negative events in the detailed application procedure as shown in Fig. 1(b). A "bank account" of these sampled negative events is added to the system and is able to store and provide the sampled negative events treated as a suppression to the probability of opposite event's occurrence. If a negative event occur, the corresponding suppression capability is transferred to the bank account. For a sampled positive event, the bank provide this "savings" which matching present positive event, then cancel the current sampling process if it is available. This bank account allows no overdraft and it does not interact with the system beyond exchanging the probability of sampled event's occurrence. Note that the additional negative event "bank" does not alter a system which is composed of large number of statistical tests.

Lots of well-developed theories will meet negative probability situation and the reason for some of them is still unknown. The most typical one is the theoretical negative cross section trouble in particle transport studies. These negative cross sections are sometime unexpectedly appeared in various theoretical approach especially when one want to determine both the position and velocity of a particle at the same time, such as the absorption and scattering cross sections for neutron inside a solid [17,18] and inelastic scattering cross sections for electron in the vicinity of a material surface [19,20]. Fortunately, presented sampling method can be employed to investigate these negative cross section, i.e. negative probabilities contribution.

In the next few paragraphs, we will focus on these theoretical negative cross section problems in electron spectroscopy techniques as an example to illustrate this presented method's strength. A quantitative understanding of electron spectroscopy techniques based on the analysis of reflected, transmitted, or emitted electrons from solid surfaces relies on an accurate description of the inelastic interaction of electrons with solids through bulk excitation and surface excitation. The inelastic interaction can be described by an important parameter, the electron inelastic mean free path (IMFP) [21]. In the bulk of a solid, the IMFP can be accurately described within the semi-classical dielectric formalism for infinite media [22], while it is position-dependent due to the spatially varying inelastic interaction when electrons are crossing an interface between two different media. A collective response of the electrons in the surface region of a solid due to the passage of electrons through the interface was predicted in the 1950s by Ritche [23] and confirmed experimentally by Powell and Swan [24]. A number of models have been developed in the last decades to calculate the quantitative information about this surface electronic excitation, i.e. position-dependent IMFP. Different approaches and approximations have been adopted: some models assume a simplified dielectric response of the solid using a classical electron-dynamics framework [23,25-28] whereas others use many-body quantum theory [29-32]. Simplifying mathematical assumptions are often made in order to highlight the relevant physics, to obtain more treatable expressions, and to keep the computation time within reasonable limits. Unfortunately, it thanks to these assumptions that all the theories available give negative IMFP values in some circumstances, for further information see [33]. Even the most mature theory within quantum mechanical formworks based on derivation of the complex inhomogeneous self-energy of the electrons is not exempt [31,32]. Fig. 2(a) show the differential inverse inelastic mean free path (DIIMFP) in the vicinity of the surface for electrons with energy of 1500 eV normal incident/emission cases from Au. The negative DIIMFP values will appear for the outgoing trajectory in a certain range of distances in the vacuum side of the sample. These negative values only appear when electron emission from the surface in the vacuum, it is smaller the further away from the surface due to the decrease in surface excitation. These negative values are more significant in the plot of inverse inelastic mean free path (IIMFP) as a function of electron depth as shown in Fig. 2(b) and its inset. The local values of IIMFP which depend on electron position can been obtained by integrating DIIMFP over the energy loss. We produced three separate IMFPs for emission electrons: traditional IIMFP only from the positive part of DIIMFP (black line); negative IIMFP only from the negative part of DIIMFP (blue line); variant IIMFP from the absolute values of DIIMFP (red line). The traditional IMFP is calculated by the popular attitude by throwing away these negative values and setting them to zero. The negative IIMFP shows significant negative values in the vacuum side, and reaches its minimum in the distance of about 12 Å. These negative cross section contribution can be estimated from the deviations between the presented IIMFP and traditional IIMFP. This deviations become larger firstly as the increasing distance when electrons moving away from surface due to the increasing share of the negative values to the total DIIMFP, and then smaller due to the decrease in surface excitation. From the percentage of negative values contribution in Fig. 2(c), these negative values in DIIMFP predominating in a few angstrom for an electron emission from a solid where the DIIMFP has a contribution only from surface excitations. From recent research [34], the so-called super-surface electron scattering, i.e. electron energy losses in vacuum above the surface of a medium, has been shown to contribute significantly to electron spectra, it is anticipated that including these negative probabilities contributions will change the transport behavior of signal electron escaping from a solid surface and give an impact on the simulated surface electron spectra but preserve more fidelity from the corresponding theory.

To prove our point, we employed the present sampling method for negative cross section problem in theoretical simulation of reflection electron energy loss spectroscopy (REELS). Fig. 3 illustrates the detail of presented sampling method works in theoretical predicting REELS spectrum by Monte Carlo method. Panel (a) shows a schematic of physical mechanism in theoretical simulation REELS spectra in presentation of Monte Carlo technique, the detail information can be found in [33]. The Monte Carlo method is able to simulate the zigzag trajectory of the electrons inside the solid as well as the energy loss of the probing electrons as a consequence of multiple inelastic scattering processes inside both the vacuum and the solid. In this study, we employ the variant IIMFP together with elastic cross section, i.e. Mott's cross section [35] to sampling the electron flight length between the successive individual scattering events, then use another random number to choose the type of scattering event according to the share of the elastic cross section, positive and negative inelastic section. In an elastic event, the scattering angle is sampled to decide the new direction of the electron movement after the collision with and atom. For a negative inelastic event, a random number produces an "opposite energy loss" from the negative part of DIIMFP, which is defined as suppression to the probability of occurring a matching energy loss process in a positive inelastic event. Instead of preforming this "opposite energy loss" immediately as an energy gain process, we store this negative event in a special "bank" for a while. When a positive inelastic event occurs, we sampling an energy loss from the positive part of DIIMFP, then take inventory of matching opposite events in this bank. If no matches, we accept this energy loss process to the current electron; otherwise, cancel this sampling results and continue the simulation. After the simulation, we can obtain the faithful spectra which take into account the negative cross section contribution. Note that, according to the present simulation model, the contribution from negative cross section was treated as a suppression to the probability of occurring energy loss event for all signal electrons instead of a simple energy gain process for a certain electron. To demonstrate the importance of the negative cross section in visible way, its contribution are estimated by tracking the canceling records of energy loss process during simulation and displayed as spectra form, i.e. negative contribution spectrum. The positive contribution spectrum can be easily obtained by deducting the negative contribution from the faithful spectra as shown in Fig. 3(d).

The simulated and measured Ag and Au REELS spectra with energy step 0.05 eV are compared in Fig. 4. The simulated spectra are normalized to and convolved with the elastic peak taken from the respective measurement. Besides faithful spectra as well as two intermediate spectra, i.e. positive and negative contribution spectra from present Monte Carlo simulation, the traditional spectra [36,37] are also presented for comparison in which the negative values are set to zero. From the negative contribution spectra, it is clear that the suppression to surface excitation caused by the negative cross section results in a reduction to energy loss peak, but enhancement to the elastic peak in the simulation. Furthermore, we also notice that the positive contribution spectra and the traditional spectra have almost the same values, and these tiny deviations origin from the slight difference of electron trajectories imply that the negative cross section weakly influence on the trajectories of signal electrons, but strongly affects the intensity of spectra directly. The deviations between faithful spectra and traditional spectra must be due to the contribution of negative cross section in present Monte Carlo simulation model. The use of presented sampling method results in REELS spectra that smaller than those from the traditional one over the whole energy range except the elastic peak, and also shift the surface excitation peak a little bit toward higher energy side. From comparisons with experimental spectra, agreement was improved from present simulation than traditional one and these improvements in comparison with measurements were amazing for the intensity of surface excitation in the low energy range, however, not significant for high energy losses due to stronger effects of multiple scattering. Significant improvement in comparison with experimental measurements verified the accuracy of the present sampling method and Monte Carlo model, since it is the first Monte Carlo simulation model including negative cross section contribution. This example is not revolutionary in the sense that although this present sampling method solved problems never solved before, but did not play a decisive role. However, this work provides a new attitudes toward the unexpected negative probability in theory that we should include these negative probability contribution in theoretical calculation at least once even without any clue, and then investigate the reasonability of these negative probability according to the compassion results.

Thoughtless neglecting these negative probabilities, we never have the opportunity to reveal the underlying mechanism hiding behind negative probability.

**ACKNOWLEDGEMENTS**

We thank Dr. S. Tanuma and Dr. H. Yoshikawa for helpful comments and discussions. This work was supported by the National Natural Science Foundation of China (Grants No. 11274288 and No. 11204289), the National Basic Research Program of China (Grants No. 2011CB932801 and No. 2012CB933702), Ministry of Education of China (Grants No. 20123402110034) and "111" project. We thank the Supercomputing Center of USTC for support in performing parallel computations.

**REFERENCES**

[1] P. A. M Dirac, Proc. Camb. Phil. Soc. **26**, 376 (1930).

[2] W. Heisenberg, Physik. Zeitschr. **32**, 737 (1931).

[3] E. P. Wigner, Phys. Rev. **40**, 749 (1932).

[4] P. A. M Dirac, Proc. Roy. Soc. London, **180**, 1 (1942).

[5] W. Pauli, Remarks on problem connected with the renormalization of quantized fields, Il Nuovo Cimento, v. 4, Suppl. No. 2, 703-710.

[6] R. P. Feynman, The Concept of Probability Theory in Quantum Mechanics, Second Berkeley Symposium on Mathematical Statistics and Probability Theory, University of California Press, Berkeley, California, pp. 553-541, 1950.

[7] R. P. Feynman, R. B. Leighton, and M. Sands, The Feynman Lectures on Physics III (Addison Wesley, Reading, MA, 1965).

[8] R. P. Feynman, Negative Probability, in Quantum Implications: Essays in Honor of David Bohm, Routledge and Kegan Paul Ltd, London and New York, pp.235-248, 1987.

[9] H. F. Hofmann, J. Phys. A: Math. Theor. **42**, 275304 (2009).

[10] J. B. Hartle, S.W. Hawking, T. Hertog, Phys. Rev. Lett. **100**, 202301 (2008).

[11] D. Sokolovski, Phys. Rev. A **76**, 042125 (2007).

[12] Y. D. Han, W. Y. Hwang, and I. G. Koh, Phys. Let. A **221**, 283 (1996).

[13] M. O. Scully, H. Walther, and W. Schleich, Phys. Rev. A **49**, 1562 (1994).

[14] T. Curtright and C. Zachos, Phys. Let. A **16**, 2381 (2001).

[15] M. Burgin, Interpretations of Negative Probabilities, Preprint arXiv:1008.1287 (2010).

[16] M. Creutz, Phys. Rev. Lett. **50**, 1411 (1983).

[17] M.B. Emmett. MORSE-CGA, A Monte Carlo Radiation Transport Code With Array Geometry Capability. ORNL-6174, America: ORNL, 1985.

[18] D. Li, Z. S. Xie, and J. M. Zhang, J. Nucl. Sci. Technol. **37**, 608 (2000).

[19] F. Salvat-Pujol, and W. S. M. Werner, Surf. Interface Anal. **45**, 873 (2013).

[20] B. Da, S. F. Mao, and Z. J. Ding, J. Phys.: Condens. Matter **23**, 395003 (2011).

[21] ISO18115 Surface Chemical Analysis—Vocabulary—Part 1: General terms and terms used in spectroscopy, International Organisation for Standardisation, Geneva, 2010.

[22] S. Tanuma, C. J. Powell, and D. R. Penn, Surf. Interface Anal. **43**, 689 (2011).

[23] R. H. Ritchie, Phys. Rev. **106**, 874 (1957).

[24] C. J. Powell, J. B. Swan, Phys. Rev. **115**, 869 (1959).

[25] E. A. Stern, and R. A. Ferrell, Phys. Rev. **120**, 130 (1960).

[26] H. Kanazawa, Prog. Theor. Phys. **26**, 851 (1961).

[27] N. Takimoto, Phys. Rev. **146**, 366 (1966).

[28] Y. C. Li, Y. H. Tu, C. M. Kwei, and C. J. Tung, Surf. Sci. **589**, 67 (2005).

[29] J. Harris and R. O. Jones, J. Phys. C: Solid State Phys. **6**, 3585 (1973).

[30] J. Harris and R. O. Jones, J. Phys. C: Solid State Phys. **7**, 3751 (1974).

[31] Z. J. Ding, J. Phys.: Condens. Matter **10**, 1733 (1998).

[32] Z. J. Ding, J. Phys.: Condens. Matter **10**, 1753 (1998).

[33] See supplemental Material at XX for details of the Monte Carlo simulation model.

[34] W. S. M. Werner, M. Novak, F. Salvat-Pujol, J. Zemek, and P. Jiricek, Phys. Rev. Lett. **110**, 086110 (2013).

[35] N. F. Mott, Proc. R. Soc. London A **124**, 425 (1929).

[36] Z. J. Ding, and R. Shimizu, Phys. Rev. B **61**, 14128 (2000).

[37] Z. J. Ding, H. M. Li, Q. R. Pu, and Z. M. Zhang, Phys. Rev. B **66**, 085411 (2002).

**FIGURE LEGENDS**

Fig. 1 A schematic diagram of present sampling method for negative probabilities. Thin solid line represent the probability density function $f(x)$ and symbol $\Omega^+$, $\Omega^-$, $|\Omega^-|$ is the positive part, negative part, absolute values of negative part, respectively. The thick solid line represent the corresponding accumulative function $F(x)$.

Fig. 2 A plot of the DIIMFP (a) for electrons of 1500 eV normal incident/emission for Au at different vertical distances measured from the surface. The corresponding traditional IIMFP, negative IIMFP and presented IIMFP as a function of the distances when electron in the vacuum side (b) and in both vacuum and solid sides (inset) are show as well as the negative IIMFP percentage in the vacuum side (c).

Fig. 3 A schematic of physical mechanism in theoretical simulation REELS spectra using Monte Carlo method (a) and the detailed information about present sampling method in this simulation including, positive inelastic event (b), negative inelastic event (c) and the simulated faithful spectra and two intermediate contribution spectra (d). Two sets of matching events are shown in (b) and (c).

Fig. 4 Comparison of the measured and simulated REELS of Ag (a) and Au (b) for 1500 eV and the angle of incidence of the primary electron beam was 50° for Ag, 35° for Au while the angle of the analyzed beam was 0° with respect to the surface normal.

1 **FIGURES**

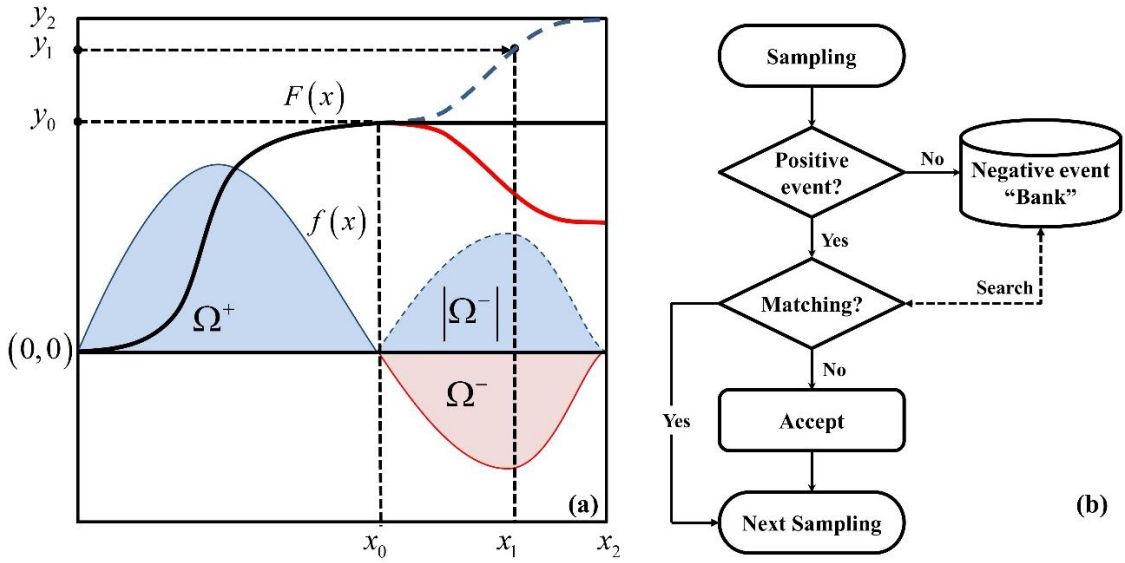

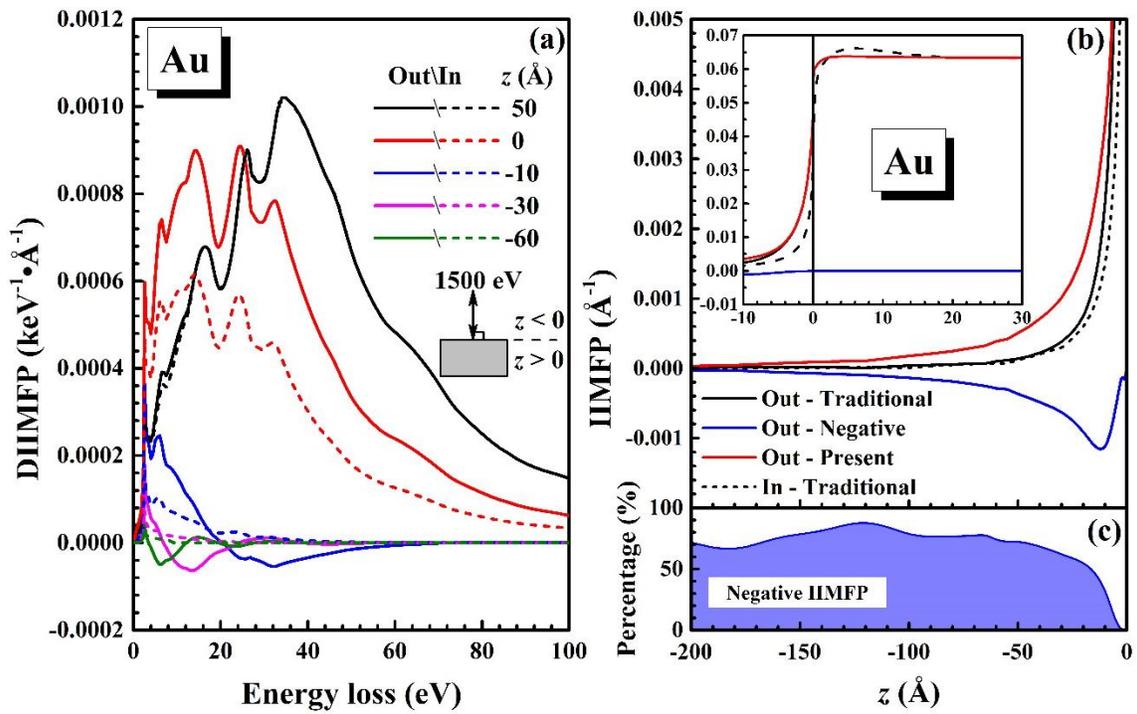

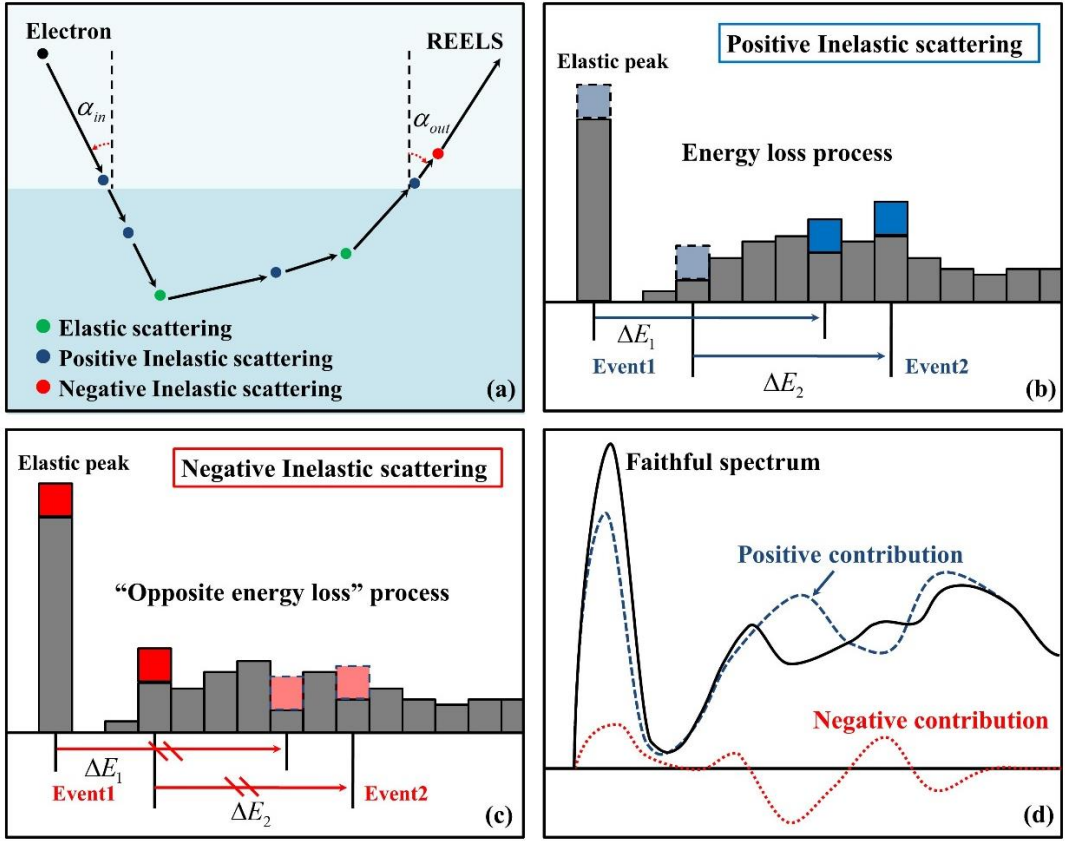

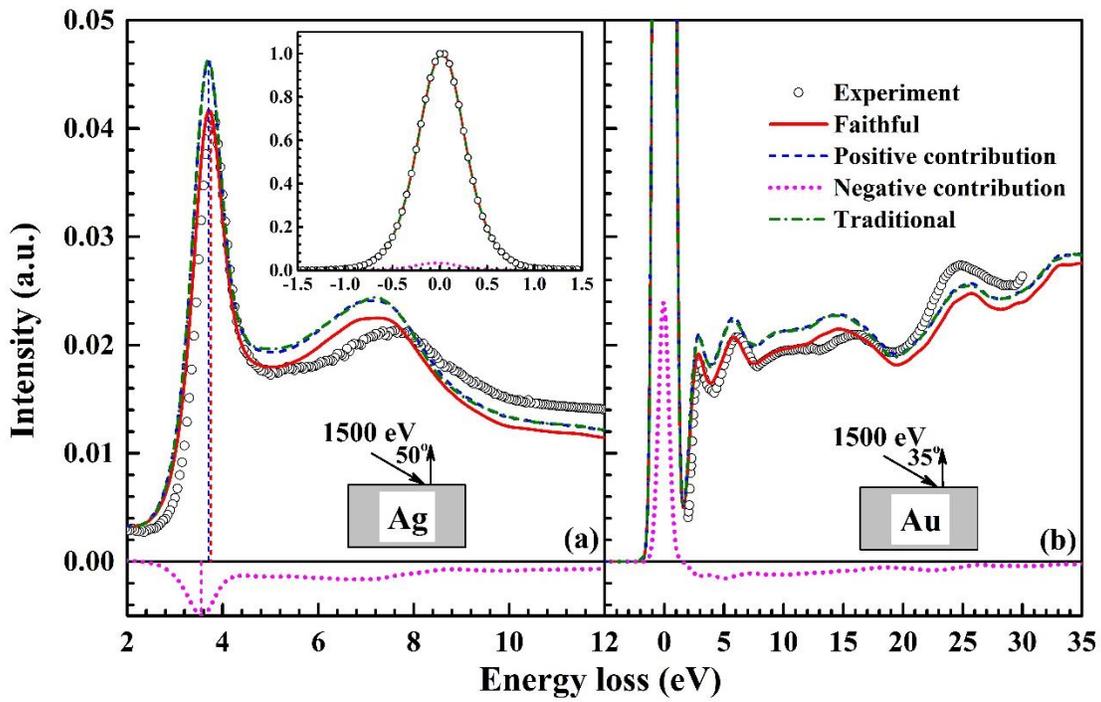

[Insert Running title of <72 characters]